\RequirePackage{lineno}
\documentclass[final,5p,times,twocolumn]{elsarticle}
\journal{Nuclear Physics A}

\usepackage{color}
\usepackage{epsfig}

\usepackage{amssymb}

\newcommand {\snn}	{\sqrt{s_{_{\rm NN}}}}
\newcommand {\gevc}	{GeV/$c$}
\newcommand {\Ncoll}	{N_{\rm coll}}
\newcommand {\pt}	{p_{T}}
\newcommand {\ptp}	{p_{T}^{\rm probe}}

\newcommand {\etap}	{\eta^{\rm probe}}
\newcommand {\phip}	{\phi^{\rm probe}}
\newcommand {\dphi}	{\Delta\phi}

\begin{document}


\title{Collisional broadening of angular correlations in a multiphase transport model}

\author[add1,add2]{Terrence Edmonds}
\author[add1,add3]{Qingfeng Li}
\author[add1,add2]{Fuqiang Wang\corref{cor}}
\cortext[cor]{Corresponding author}
\ead{fqwang@zjhu.edu.cn}
\address[add1]{School of Science, Huzhou University, Huzhou, Zhejiang 313000, P.R.~China}
\address[add3]{Institute of Modern Physics, Chinese Academy of Sciences, Lanzhou, Gansu 730000, P.R.~China}
\address[add2]{Department of Physics and Astronomy, Purdue University, West Lafayette, Indiana 47907, USA}

\begin{abstract}
Systematic comparisons of jetlike correlation data to radiative and collisional energy loss model calculations are essential to extract transport properties of the quark-gluon medium created in relativistic heavy ion collisions. This paper presents a transport study of collisional broadening of jetlike correlations, 
by following parton-parton collision history in a multiphase transport (AMPT) model. The correlation shape is studied as functions of 
the number of parton-parton collisions suffered by a high transverse momentum probe parton ($\Ncoll$) and the azimuth of the probe relative to the reaction plane ($\phip_{\rm fin.}$). Correlation is found to broaden with increasing $\Ncoll$ and $\phip_{\rm fin.}$ from in- to out-of-plane direction. This study provides a transport model reference for future jet-medium interaction studies.
\end{abstract}

\begin{keyword}
collisional energy loss \sep parton transport \sep jetlike correlations

\PACS 25.75.-q \sep 25.75.Gz

\end{keyword}

\maketitle

\section{Introduction}
Relativistic heavy ion collisions aim to create a quark-gluon plasma (QGP) to allow studies of quantum chromodynamics (QCD) at the extreme conditions of high temperature and energy density~\cite{Arsene:2004fa,Back:2004je,Adams:2005dq,Adcox:2004mh,Muller:2012zq}. The system created in these collisions is described well by hydrodynamics where the high pressure buildup drives the system to expand at relativistic speed~\cite{Heinz:2013th,Gale:2013da}. Hydrodynamics inspired model fit to experimental data at low transverse momentum ($\pt$) is consistent with particles being locally thermalized with a common radial flow velocity~\cite{Abelev:2008ab}. On the other hand, high-$\pt$ particles (partons) are produced by hard (large momentum transfer) processes, and emerge in the final state as a collimated shower of hadrons (jets)~\cite{Jacobs:2004qv}. They traverse the medium and lose energy as they interact with it~\cite{Gyulassy:1990ye,Wang:1991xy,Jacobs:2004qv}. Due to the energy loss, the number of high-$\pt$ particles is suppressed by a factor of several in heavy ion collisions compared to properly normalized proton-proton collisions~\cite{Adcox:2001jp,Adler:2002xw,Adler:2003qi,Adams:2003kv,Adams:2003im,Adler:2003ii,Aad:2010bu,Aamodt:2010jd,Aamodt:2011vg,Chatrchyan:2011sx,Aad:2012vca,Chatrchyan:2012nia,CMS:2012aa,Chatrchyan:2013kwa,Adam:2015doa}. How the parton showers (which are reflected in jet shapes) change from proton-proton collisions to heavy ion collisions, due to parton-medium interactions, is a valuable tool to study jet-medium interactions~\cite{Wang:2013qca}. 

Jets lose energy via gluon radiations
 and collisions of the parent parton and/or the daughter fragments with medium particles~\cite{Baier:2000mf}. Particle angular correlations with high-$\pt$ particles, originating from jets, contain all those information, including the medium particles that are not originated from jets but now correlated with the jet due to collisional energy loss~\cite{Wang:2013qca}. Collisional energy loss has been studied by various models~\cite{Mustafa:2003vh,Djordjevic:2006tw,Adil:2006ei,Qin:2007rn,Peigne:2008nd,Shin:2010hu,Berrehrah:2014kba}. 
The focus of those studies has been on the amount of energy loss by energetic partons, not on the angular correlations induced by parton-parton collisions. In this paper, we study the effect of collisional energy loss on jetlike correlations. We use, for the first time, a multiphase transport (AMPT) model~\cite{Zhang:1999bd,Lin:2001zk,Lin:2004en} for our study. We trace the collision history~\cite{He:2015hfa} of high-$\pt$ partons (referred to as probe partons), one by one, with partons from the medium (referred to as medium partons). We study the angular correlation width as a function of 
the number of parton-parton collisions suffered by each high-$\pt$ probe parton ($\Ncoll$). Of particular interest are non-central collisions where the overlap zone of the colliding nuclei is anisotropic in the transverse plane (perpendicular to beam)~\cite{Ollitrault:1992bk}. The interaction strength varies with the azimuthal angle of the probe parton relative to the reaction plane (RP), which can be experimentally accessible via the event plane~\cite{Agakishiev:2010ur,Agakishiev:2014ada}. 

\section{Model Setup and Analysis Technique}
AMPT~\cite{Zhang:1999bd,Lin:2001zk,Lin:2004en} has been widely used to describe experimental data. The string melting version of AMPT~\cite{Lin:2001zk,Lin:2004en} reasonably reproduces particle yields, $\pt$ spectra, and elliptic flow of low-$\pt$ pions and kaons in central and mid-central Au+Au collisions at nucleon-nucleon center-of-mass energy of $\snn=200$~GeV and Pb+Pb collisions at $\snn=2760$~GeV~\cite{Lin:2004en,Xu:2011fi,Xu:2011jm,Lin:2014tya,Ma:2016fve,He:2017tla}.
We employ the same string melting version of AMPT (v2.26t5) 
in this study. The model consists of fluctuating initial conditions inherited from Glauber nuclear geometry in HIJING~\cite{Wang:1991hta,Gyulassy:1994ew}, $2\to2$ parton elastic scatterings~\cite{Zhang:1997ej}, quark coalescence for hadronization~\cite{Lin:2004en,Lin:2009tk}, and hadronic interactions~\cite{Li:1995pra,Lin:2004en}. Parton-parton interactions stop when no parton pairs can be found within their interaction range, given by the parton-parton interaction cross section. We use Debye screened differential cross-section, $d\sigma/dt\propto\alpha_s^2/(t-\mu_D^2)^2$~\cite{Lin:2004en}, with strong coupling constant $\alpha_s=0.33$ and Debye screening mass $\mu_D=2.265$/fm. The total parton scattering cross section is then $\sigma=3$~mb. 

We note, however, AMPT does not contain jets; it is a model used mainly to study low-$\pt$ physics. AMPT uses HIJING as its initial condition. The final-state hadrons generated by HIJING are decomposed into their constituent quarks, uniformly in azimuth in the hadron rest frame~\cite{private}. The individual quarks are then allowed to scatter elastically until partonic freezeout, i.e.~hadronization. Thus, although (di)jets are produced initially in HIJING simulation, the angular correlations among the hadrons initially from HIJING are mostly destroyed. The quarks liberated from those hadrons form new hadrons after rescatterings that are no longer those same hadrons from HIJING. 
It is, however, still relevant to use a high-$\pt$ parton in AMPT as a probe and study the effect of its collisional energy loss as it propagates through the medium. 
In this work we simply define initial partons with $\pt>3$~\gevc\ as our definition of initial `jet', or probe particle.
To have focused discussion, we will use the jet terminology, namely jetlike correlations, to refer to the parton-parton correlations we study in this paper. 

We have simulated $10^5$ Au+Au collisions with fixed impact parameter $b=8$~fm at $\snn=200$~GeV using AMPT. 
We trace the collision history of the initially produced high-$\pt$ probe partons. We know exactly which medium partons have interacted with the probe partons. We study the correlations of those medium partons with the probe partons. We focus only on the parton level angular correlations. All results shown in this study are for partons within pseudo-rapidity window $|\eta|<2$; $\eta\equiv-\ln\theta/2$ where $\theta$ is the polar angle relative to the beam direction.

\section{Results}
\subsection{Collisional broadening}
\begin{figure}[hbt]
  \begin{center}
    \includegraphics[width=0.4\textwidth]{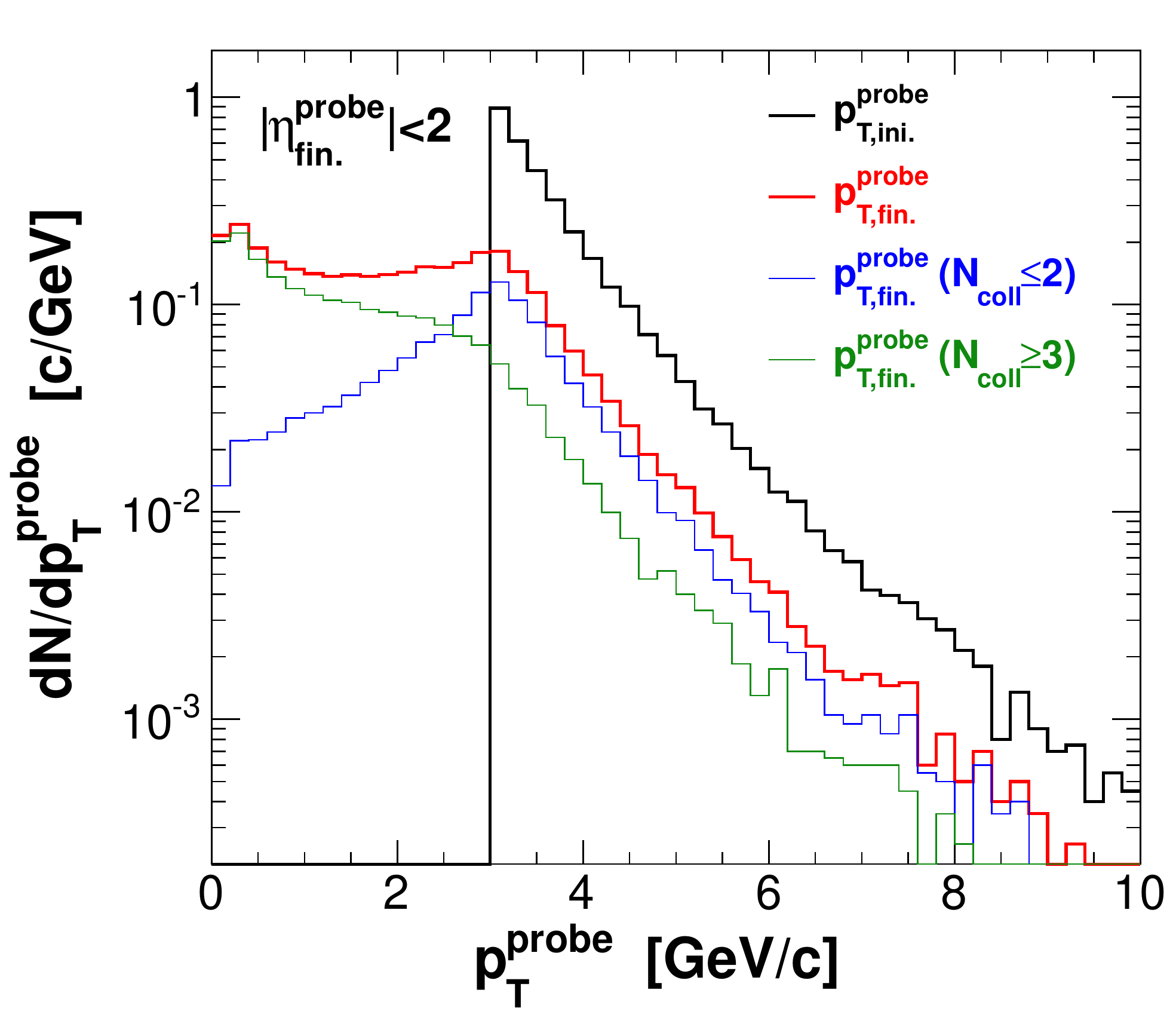}
    \caption{(Color online) Probe parton $\ptp$ distributions in $b=8$~fm Au+Au collisions at $\snn=200$~GeV in AMPT: initial $\pt$ (thick black), final $\pt$ (thick red), final $\pt$ for probe partons with $\Ncoll\leq2$ (thin blue) and $\Ncoll\geq3$ (thin green). Partons are required to have final-state $|\etap_{\rm fin.}|<2$.}
    \label{fig:tstPt}
  \end{center}
\end{figure}

Figure~\ref{fig:tstPt} shows the $\pt$ distributions of the probe partons. The black histogram shows the distribution of the initial probe parton $\pt$ (which we require to be at least 3~\gevc). The red histogram shows the final-state $\pt$ distributions of those partons. Significant energies are lost on average by the high-$\pt$ probe partons and the final-state $\pt$ can be very small. The number of high-$\pt$ partons is reduced by a factor of 3-4, and the reduction factor is approximately $\pt$ independent. The blue and green histograms show the final-state $\pt$ distributions of partons that have suffered, respectively, no more than two and more than two collisions with other partons. The more the collisions, the more the energy loss on average. Due to energy loss (parton-parton interactions in general), partons emerging in one particular direction come preferentially from the region close to the surface in that direction. This surface bias is stronger for higher $\pt$ partons emerging in the final state.

\begin{figure*}[hbt]
  \begin{center}
    \includegraphics[width=0.325\textwidth]{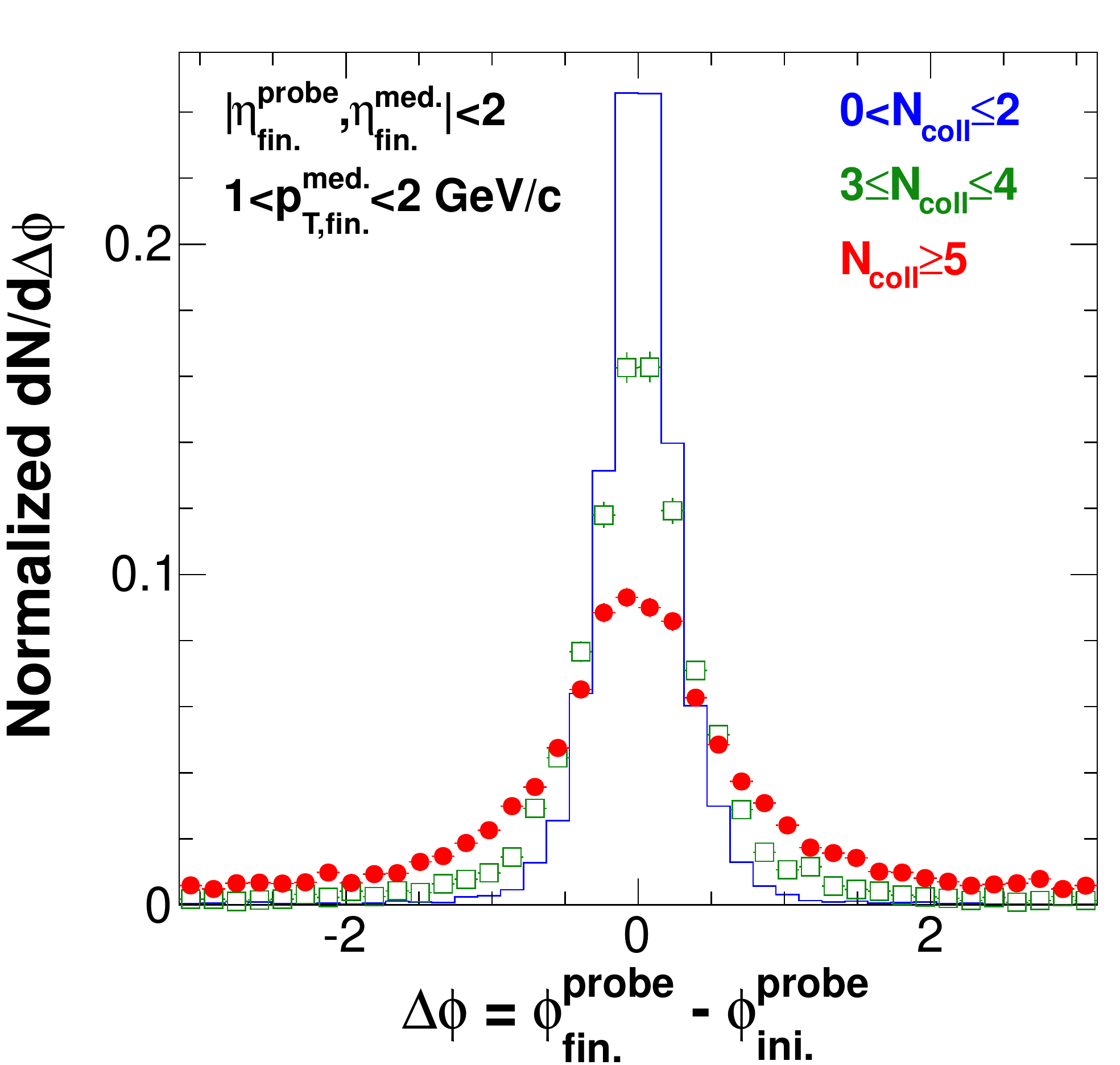}
    \includegraphics[width=0.325\textwidth]{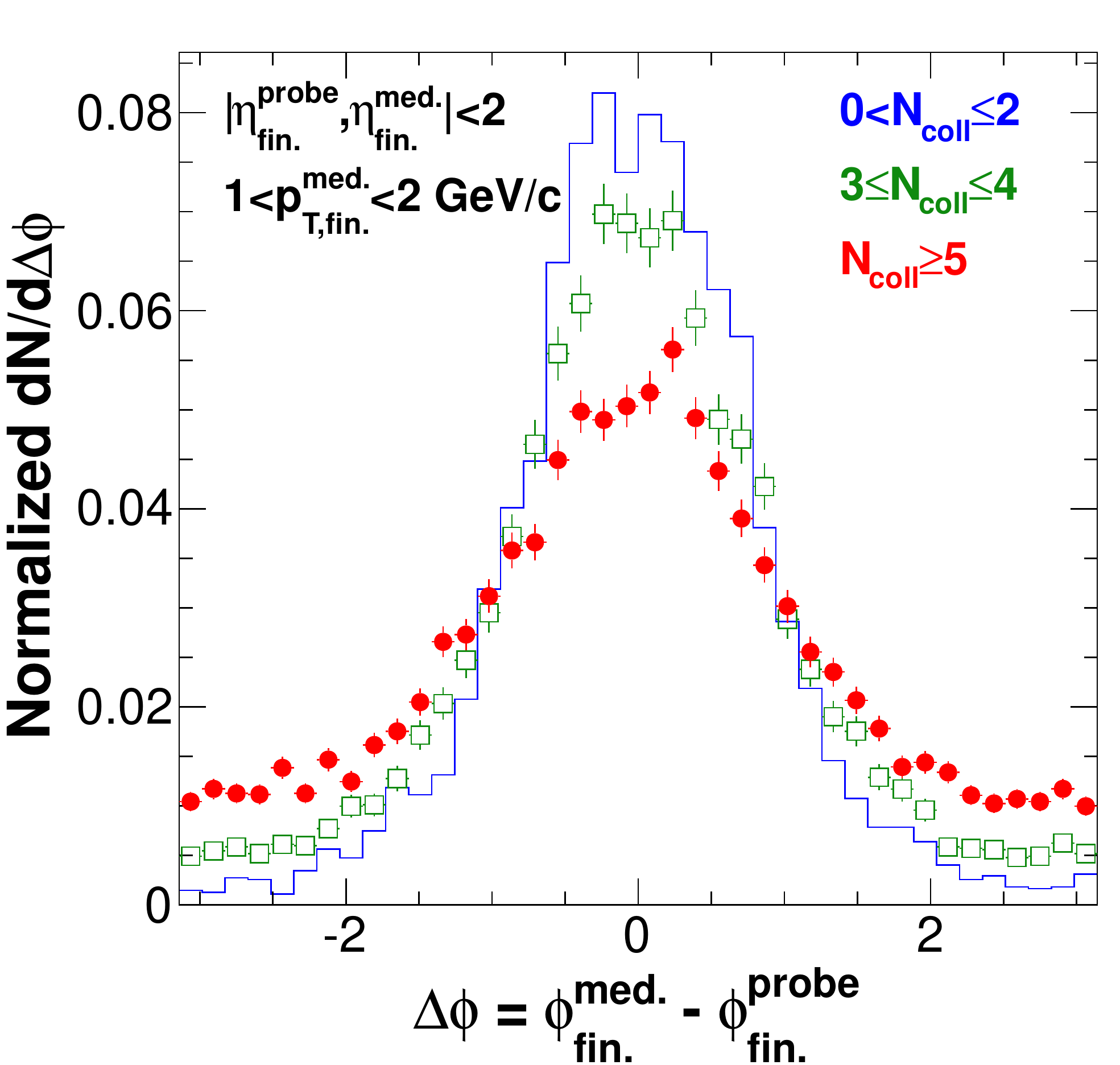}
    \includegraphics[width=0.325\textwidth]{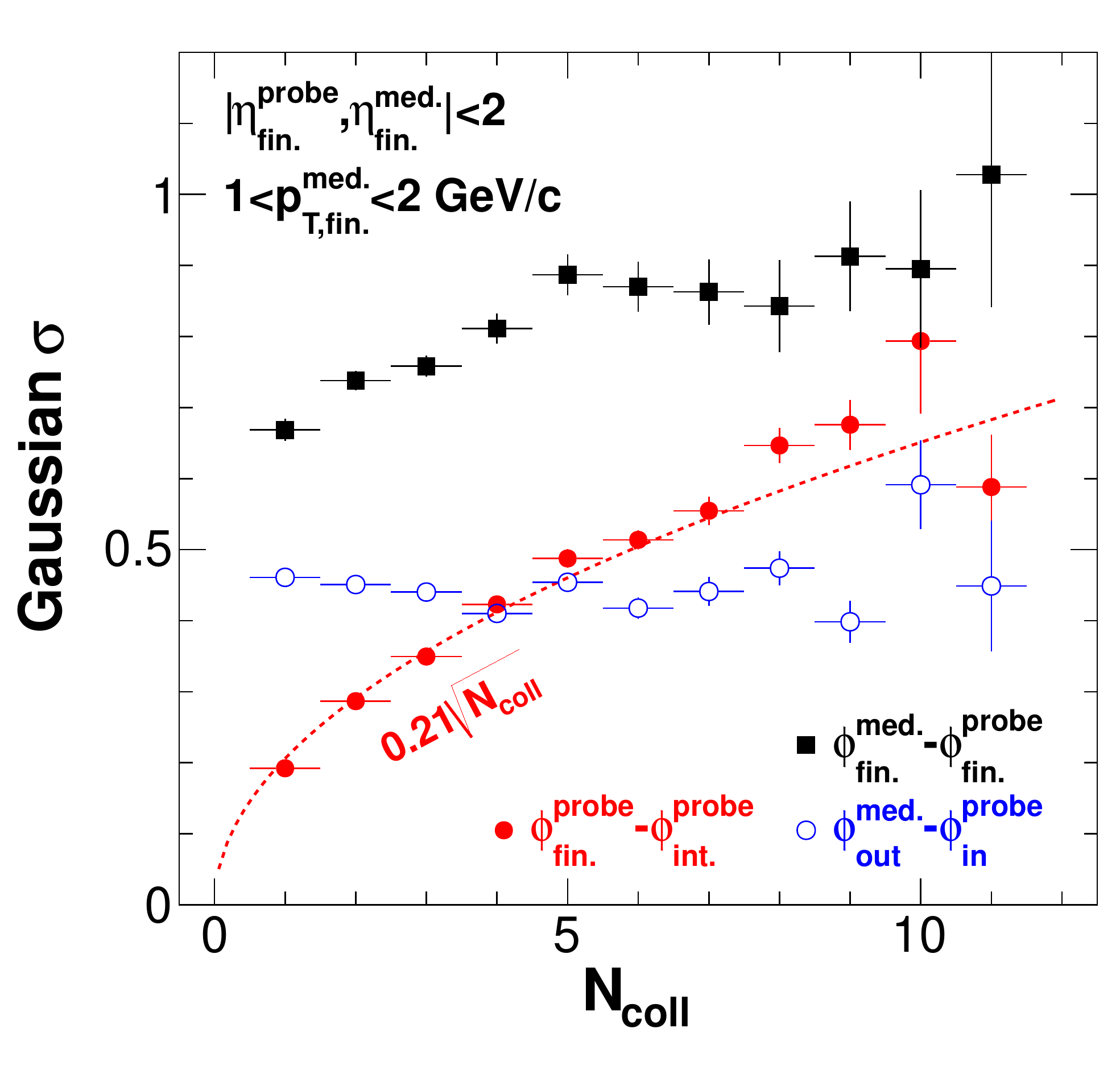}
    \caption{(Color online) Azimuthal dispersions of probe partons (left), azimuthal correlations between medium and probe partons (middle) for different $\Ncoll$ suffered by the probe parton, and the Gaussian+pedestal (Eq.~(\ref{eq})) fit $\sigma$ as a function of $\Ncoll$ (right) in $b=8$~fm Au+Au collisions at $\snn=200$~GeV in AMPT. The probe and medium partons are within $|\eta_{\rm fin.}|<2$ in the final state. The medium parton final-state $\pt$ range is $1<p_{T,{\rm fin.}}^{\rm med.}<2$~\gevc\ and no cut is applied on the probe parton $\pt$.}
    \label{fig:dphi_sigma_vs_Ncoll}
  \end{center}
\end{figure*}

The probe parton changes direction at each collision with a medium parton. The more the collisions, the larger the angular dispersion. Figure~\ref{fig:dphi_sigma_vs_Ncoll} (left panel) shows the normalized distribution in the probe parton azimuthal angle change ($\dphi$) from initial to final state for different $\Ncoll$ values. The angular dispersion increases with increasing $\Ncoll$. The parton initial direction is, however, inaccessible experimentally. One has to rely on final state particle correlations to access this physics. This is shown in Fig.~\ref{fig:dphi_sigma_vs_Ncoll} (middle panel) where the correlation functions in azimuthal angle difference ($\dphi$) between medium parton and probe parton, both at final state~\footnote{Although the final state parton information is also experimentally inaccessible, they better reflect the final-state hadrons, modulo hadronization and hadronic rescattering effects.}, are depicted for different $\Ncoll$ ranges. The correlation broadens with increasing $\Ncoll$ as expected. The medium partons included in the plot are only those that have interacted with the probe parton in the evolution history. We do not need to do any background subtraction in this study because we know exactly which partons from the medium have interacted with the probe parton. However, the medium partons that have interacted with the probe parton can suffer subsequently collisions with other medium partons, and those other medium partons may indirectly become correlated with the probe parton. We do not include those secondary medium partons in our present study. 

To study the broadening more quantitatively, we fit the $\dphi$ distributions in both the left and middle panels of Fig.~\ref{fig:dphi_sigma_vs_Ncoll} to the sum of a Gaussian plus a constant pedestal (referred to hereon as Gaussian+pedestal), 
\begin{equation}
\frac{dN}{d\dphi}=A\exp\left(-\frac{(\dphi-\alpha)^2}{2\sigma^2}\right)+C\,,
\label{eq}
\end{equation}
where $A, \sigma,$ and $C$ are fit parameters. The $\alpha$ parameter is fixed to zero as required by symmetry; fits treating $\alpha$ as a free parameter yield fit results consistent with zero.
The Gaussian $\sigma$ is shown in the right panel of Fig.~\ref{fig:dphi_sigma_vs_Ncoll} as a function of $\Ncoll$. The probe parton dispersion can be approximately described by $0.21\sqrt{\Ncoll}$; the dispersion is about $12^{\rm o}$ per collision. The broadening in medium-probe correlation is primarily due to the angular dispersion of the probe partons. This is because the individual medium parton that interacts with the probe parton should not know about how many collisions the probe parton has suffered, except the probe parton $\pt$ is degraded with $\Ncoll$. This point is illustrated by the open circles where the angular width of the outgoing medium parton relative to the incoming probe parton at each collision, integrated over all such medium partons, is shown as a function of $\Ncoll$. No significant dependence is observed over the $\Ncoll$ range shown in Fig.~\ref{fig:dphi_sigma_vs_Ncoll} right panel.

The above results are restricted to a particular medium parton $\pt$ range. We have also studied the correlations as a function of the medium parton $\pt$. The correlation function is broader for lower $\pt$ medium partons. This is expected because lower $\pt$ particles can be more easily scattered to large angles. 

\subsection{Pathlength dependence}
\begin{figure}[hbt]
  \begin{center}
    \includegraphics[width=0.4\textwidth]{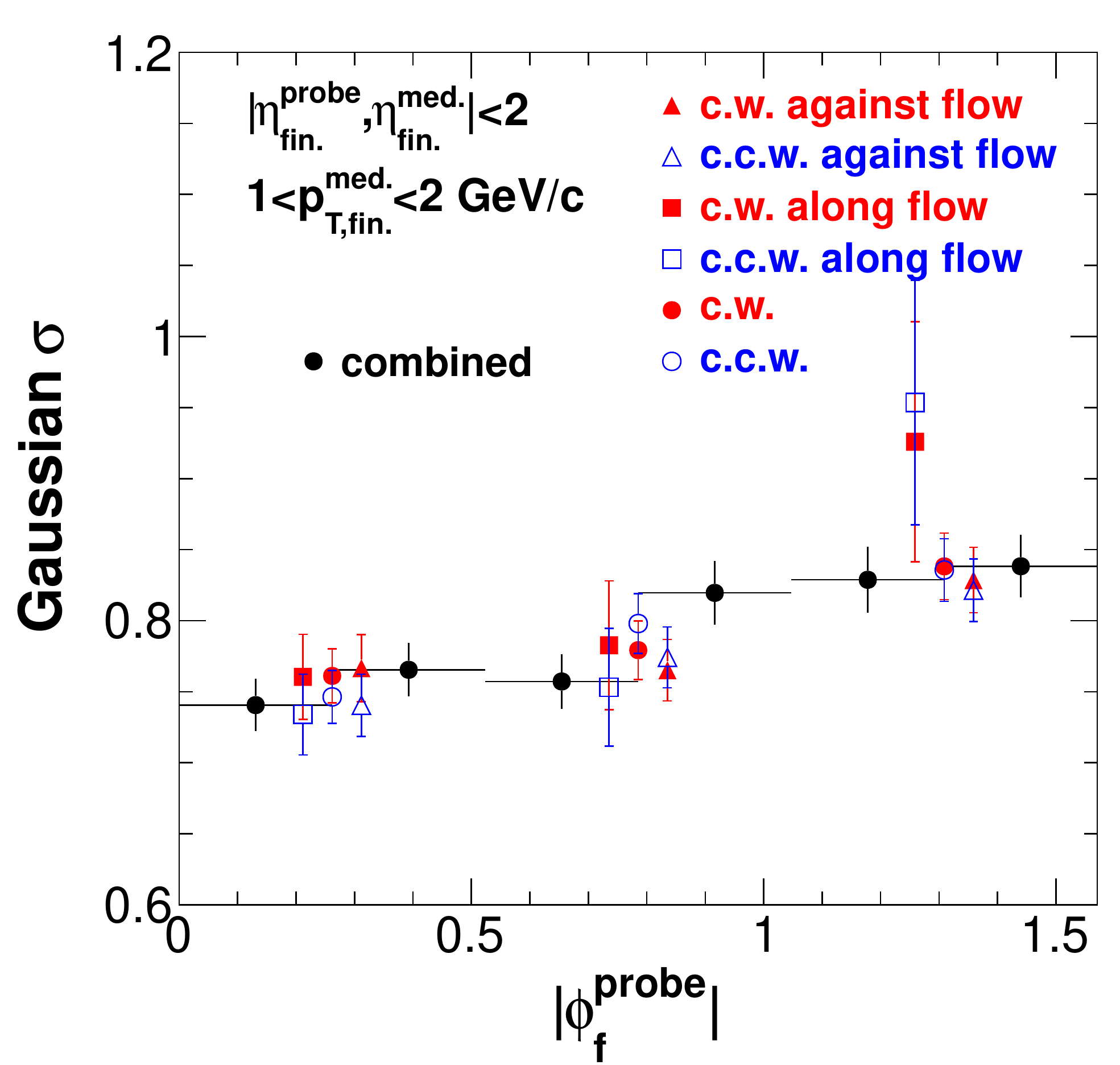}
    \caption{(Color online) The Gaussian+pedestal (Eq.~(\ref{eq})) fit $\sigma$ as a function of the probe parton $\phip_{\rm fin.}$ (filled black circles) in $b=8$~fm Au+Au collisions at $\snn=200$~GeV in AMPT. The other, colored data points show the corresponding fit $\sigma$ for the different configurations studied in Fig.~\ref{fig:dphi_asym_tstPhi}; some of the data points are shifted horizontally for clarity.}
    \label{fig:dphi_asym_tstPhi_width}
  \end{center}
\end{figure}

In non-central collisions, the overlap collision zone between the two nuclei is almond shaped. One expects partons in the long-axis direction to suffer more collisions than those in the short-axis direction (the reaction plane direction). This is indeed true in AMPT. As a consequence, the correlation function broadens as the probe parton varies from the short-axis direction to the long-axis direction. This is shown in Fig.~\ref{fig:dphi_asym_tstPhi_width} by the filled black circles where the Gaussian+pedestal (Eq.~(\ref{eq})) fit $\sigma$ is plotted as a function of the probe parton azimuthal angle relative to the reaction plane, $\phip_{\rm fin.}$. In this study, we take the reaction plane direction as known in AMPT, fixed at zero radian. The other data points will be discussed later. 

\begin{figure*}[hbt]
  \begin{center}
    \includegraphics[width=0.325\textwidth]{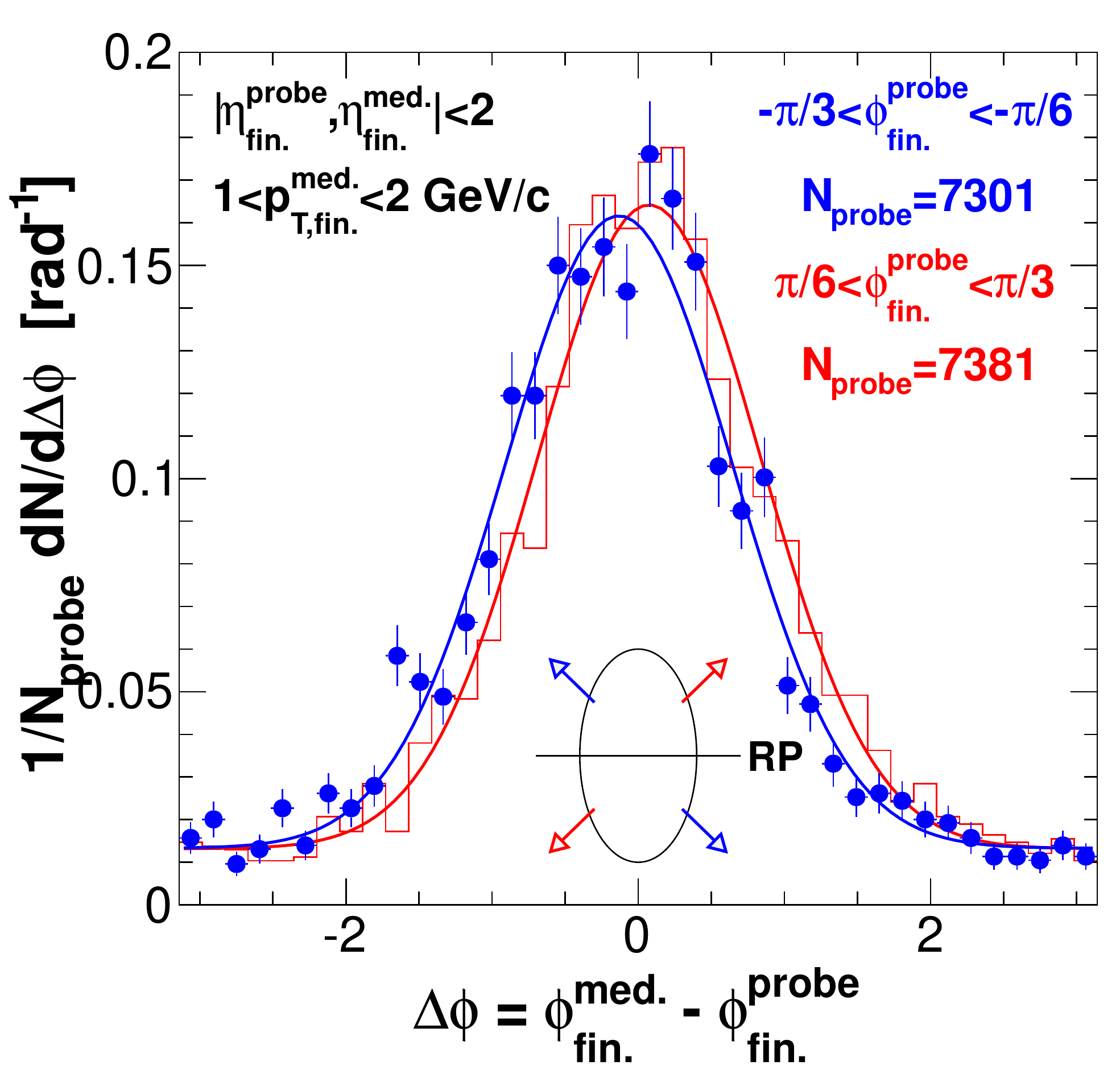}
    \includegraphics[width=0.325\textwidth]{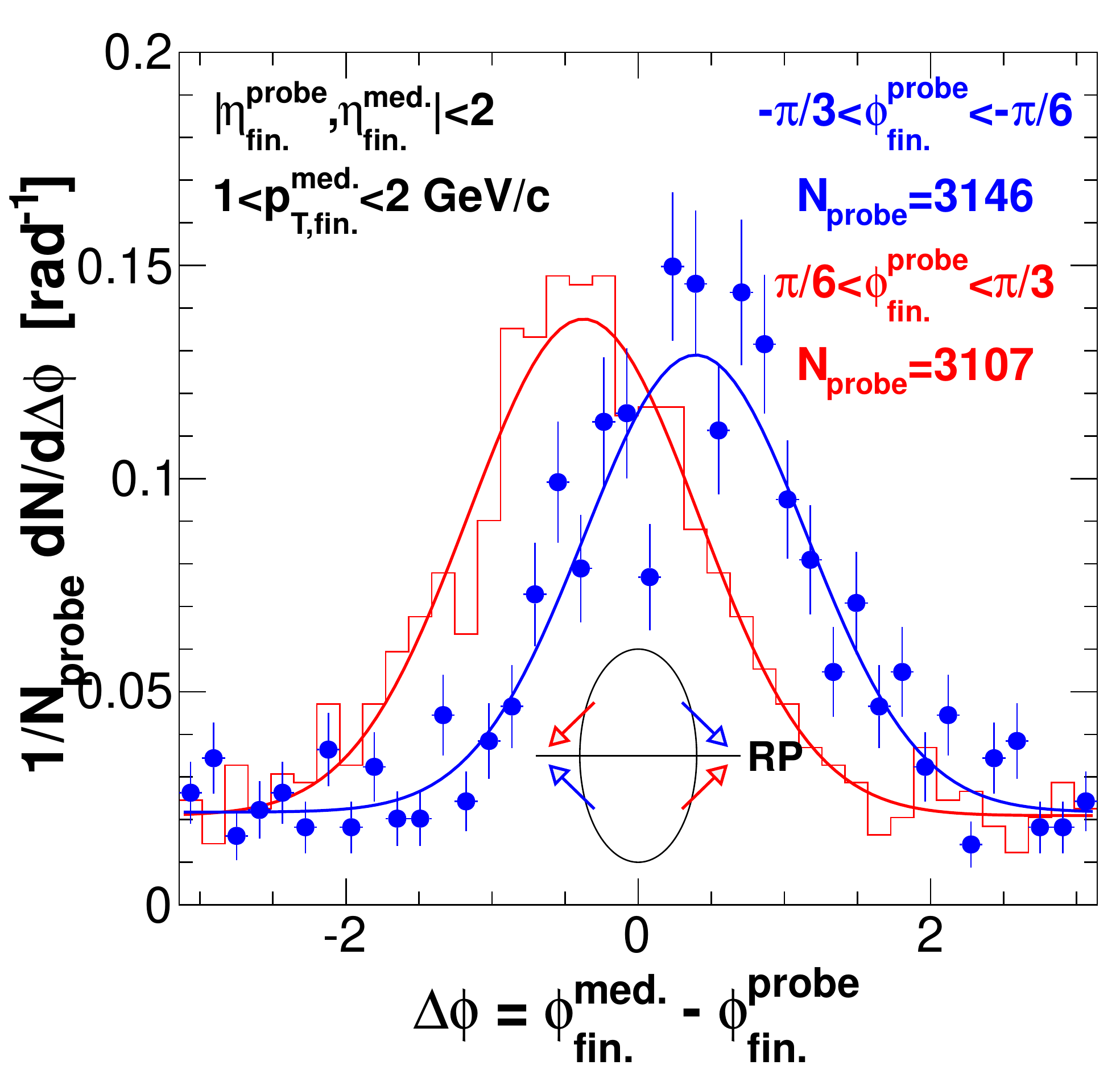}
    \includegraphics[width=0.325\textwidth]{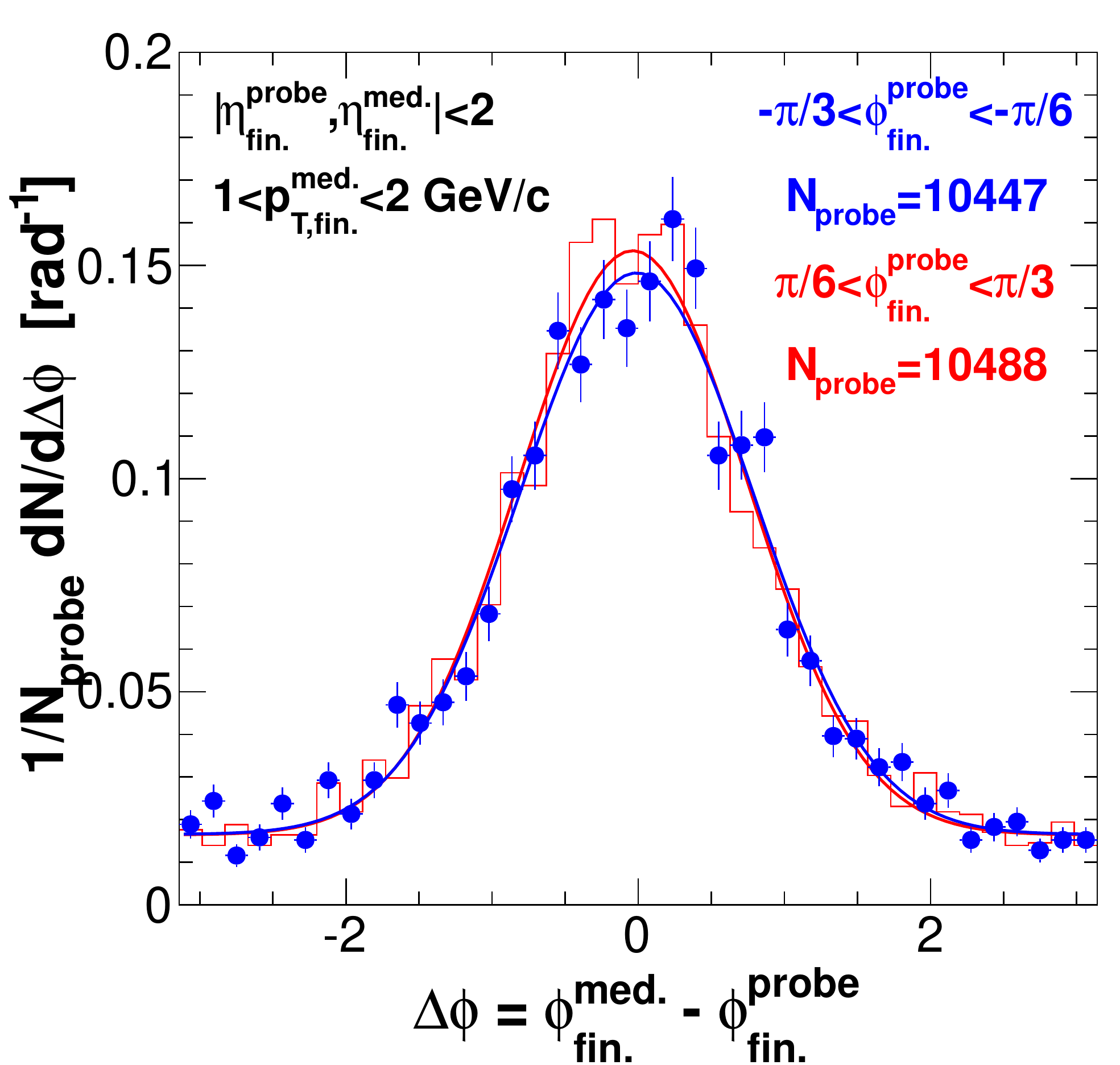}
    \caption{(Color online) Per probe normalized azimuthal correlation functions in $b=8$~fm Au+Au collisions at $\snn=200$~GeV in AMPT, for probe partons within $\pi/6<\phip_{\rm fin.}<\pi/3$ (red histograms) and $-\pi/3<\phip_{\rm fin.}<-\pi/6$ (blue points), separately. (left) The probe parton moves ``along'' the medium flow; (right) The probe parton moves ``against'' the medium flow; (c) The sum of the two cases. See the insert cartoons.}
    \label{fig:dphi_asym_tstPhi}
  \end{center}
\end{figure*}

Because of parton-parton interactions, the parton's final-state momentum is generally correlated with its freeze-out position. This is referred to as radial flow and could be a consequence of hydrodynamic expansion of the collision system. The probe partons generally move along with the medium partons. For example, the majority of probe partons with positive $\phip_{\rm fin.}$, i.e.~counterclockwise (c.c.w.) from the RP, comes from the first (and third) quadrant. However, there are probe partons with positive $\phip_{\rm fin.}$ that come from the fourth (and second) quadrant, although their population is relatively small. These partons move more or less orthogonally with respect to the medium flow. Similarly for probe partons with negative $\phip_{\rm fin.}$, i.e.~clockwise (c.w.) from the RP. One can imagine that the correlation functions of these two groups of probe partons with the medium partons would be very different. This is indeed true and shown in Fig.~\ref{fig:dphi_asym_tstPhi} comparatively between the left and middle panels; the direction of motion of the probe is indicated by the arrows of the corresponding color (color online) in the cartoon insert. For those probe partons moving along the medium flow (left panel), the fit Gaussian centroids are slightly offset from $\dphi=0$. 
For those probe partons that move orthogonally from the medium flow (middle panel), the angular correlation function is significantly offset from $\dphi=0$; which direction of the offset depends on whether the probe parton is c.w.~or c.c.w. For probe partons moving in the direction c.c.w.~away from the RP (red arrows in the insert cartoon in Fig.~\ref{fig:dphi_asym_tstPhi} middle panel), the correlation is strongly shifted toward negative $\dphi$, because the medium partons mostly move in the direction toward the more negative azimuth from the probe parton direction. Likewise, for those moving in the direction c.w.~away from the RP (blue arrows in the insert cartoon in Fig.~\ref{fig:dphi_asym_tstPhi} middle panel), the correlation is shifted to the positive side. 
Note the correlation functions for probes c.w.~and c.c.w.~away from the RP are, by definition, reflectively symmetric about $\dphi=0$, and this is apparent in Fig.~\ref{fig:dphi_asym_tstPhi}.
The correlation functions have been normalized, divided by the number of probe partons ($N_{\rm probe}$). There are relatively fewer probe partons that move against than along the medium flow, and this is reflected by the relative statistical precisions between the left and middle panels of Fig.~\ref{fig:dphi_asym_tstPhi}.

Unfortunately, because the parton configuration space positions are not experimental accessible, one cannot select probes according to the directions in Fig.~\ref{fig:dphi_asym_tstPhi} left and middle panels. One can only select particles according to their momentum direction, either c.w.~or c.c.w.~away from the RP. The correlation functions for such selections are shown in the right panel of Fig.~\ref{fig:dphi_asym_tstPhi}. The offsets appear to be averaged out (note the average is weighted by the number of probe partons), and the correlation functions do not seem to be asymmetric any more, according to AMPT. This result is surprising as the sum correlation function does not have to be symmetric a priori; there is no physics requirement for the symmetry in the correlation function. The shape is determined by the relative abundance of the probe parton populations and the shapes of the individual correlation functions of the two sources. The fact that it is symmetric from AMPT must be accidental. Experimentally, it is still interesting to look for this possible asymmetry as it would provide sensitive information to the medium dynamics. 

The data in Fig.~\ref{fig:dphi_sigma_vs_Ncoll} and the combined data in Fig.~\ref{fig:dphi_asym_tstPhi_width} do not distinguish between c.w.~and c.c.w.~directions or between configurations along and against flow. 
Since the correlation functions are the sum of various sources, it is important to ask whether the broadening from in- to out-of-plane observed in Fig.~\ref{fig:dphi_asym_tstPhi_width} is due to real dynamics or simply different weightings of the various sources for in- and out-of-plane probe partons. 
To check this, we show in Fig.~\ref{fig:dphi_asym_tstPhi_width} also the Gaussian fit $\sigma$ for the various selections of the probe parton direction, together with that of the overall correlation function without distinguishing between c.w.~and c.c.w.~directions in the filled black circles. 
The general increasing trend from in-plane to out-of-plane is observed for all selections. This indicates that the broadening is not simply due to the asymmetric correlation functions integrated together, but also physics dynamics, namely the pathlength dependent effect of parton-parton interactions.

Experimental studies of jetlike correlations at low-to-intermediate $\pt$ suffer from large uncertainties in flow background subtraction~\cite{Adams:2005ph,Adler:2005ee,Adare:2008ae,Aggarwal:2010rf,Agakishiev:2010ur,Agakishiev:2014ada}. Recent studies, with a novel, robust flow subtraction method, show that jetlike correlation width increases with collision centrality (i.e.~amount of jet-medium interactions)~\cite{Jiang:2015kva}. The reaction-plane dependent correlation studies with the new flow subtraction method are not yet available, but would be important in illustrating pathlength dependent effects of jet-medium interactions.

\section{Summary}
We have studied the effect of collisional energy loss on jetlike correlations using, for the first time, the AMPT parton transport model. We follow the parton cascade history in AMPT to identify the partons that have interacted with the probe partons, so that we do not include the underlying event in our correlation functions and no background subtraction is needed. 
We found that the jetlike correlations broaden with increasing number of parton-parton interactions suffered by the probe parton, and with probe parton azimuthal angle from in-plane direction to out-of-plane direction. This broadening is an indication of the pathlength dependent effect from interactions between probe partons and the medium. In addition, there are dramatically asymmetric correlation functions in AMPT depending on the relative orientations of the probe parton with respect to the medium radial flow. These asymmetric correlation functions are averaged out in AMPT in the final ``measured'' one. Symmetric correlation functions measured in experiments would thus by no means be any indication of weak effects of medium flow on jetlike correlations.

Jetlike correlation shape modification in heavy ion medium contains valuable information about the medium properties and partonic energy loss mechanisms, including both radiative and collisional energy losses. 
In order to learn about the medium properties, comparisons of jetlike correlation data to combinations of various radiative and collisional energy loss models are essential. 
The AMPT model contains only elastic collisional energy loss; radiative energy loss effects cannot be studied with AMPT. 
Nevertheless, our transport model study of collisional energy loss effect on jetlike correlations provides a proof of concept that angular correlations are broadened by collisional energy loss, and thus should be useful to the study of jet-medium interactions. It would be valuable in the future to compare collisional angular broadening from AMPT to those from collisional as well as radiative energy loss models.

As mentioned in the introduction, there is no dijet production in AMPT. We have only studied the correlations with respect to a single, initially energetic parton produced in AMPT. In the future, we plan to study collisional energy loss and jet broadening by embedding Pythia dijets into AMPT and follow the collision history of both jet partners. 

\section{Acknowledgments}
T.E.~and F.W.~thank Zi-Wei Lin and Liang He for discussions and help on the AMPT model. This work was supported in part by the National Natural Science Foundation of China (Grant Nos. 11375062, 11647306) and the U.S.~Department of Energy (Grant No.~DE-FG02-88ER40412). 

\bibliographystyle{elsarticle-num} 
\bibliography{../../ref}
\end{document}